\documentclass[12pt]{article}
\usepackage{amsfonts}

\usepackage{amsmath}
\usepackage{floatflt}
\usepackage{graphicx}
\usepackage{cite}

\begin{document}

\title{Nucleation of $p$-branes with form fields and dielectric brane}
\author{Pei Wang \thanks{%
peiwang@nwu.edu.cn} ~and Hua Jiang \\
Institute of Modern Physics, Northwest University, Xi'an, 710069, China}
\date{}
\maketitle

\begin{abstract}
In this paper we discuss how to generalize the concept of nucleation to the $%
p$-branes with form fields. And we try to get ready for calculating the
decay width of the dielectric brane.
\end{abstract}


\section{Introduction}

$P$-branes, including of M-branes, D branes and NS-NS branes, in string/M
theory are very important objects which appear in almost every recent paper
in this area. The earlier references are reviewed in \cite{1,2,3,4}. Several
authors have given the static solutions of branes, see e.g. \cite{2,5,6,7,8}%
. Moreover, basing on the higher dimensional rotating black holes \cite{9},
Cveti\v{c} and Youm have also developed the rotating M-branes \cite{10,11}.

In another respect, a new kind of branes, called flux branes, which are
generalized from 4-dimensional Melvin universe \cite{12} are researched in
many papers \cite{13,14,15,16,17,18,19,20}. Among other things the authors
of ref. \cite{21} and \cite{22} apply the flux-brane technique to the
M-theory and find the supergravity solutions describing certain dielectric
branes in 10 dimensions \cite{23}. In ref. \cite{24} and \cite{21} the
authors compare the supergravity solution with Myers' dielectric brane, the
worldvolume theory, and find them in agreement with each other. Recently a
paper appeared, in which the decay of dielectric brane is discussed by using
the open string technique \cite{25}. To look for its supergravity
counterpart, we have to study the nucleation of dielectric branes. This
means that before we consider the dielectric brane we should find the
nucleation of $p$-branes first. That is the goal of this paper. Dowker et
al. already studied these questions in a remarkable paper \cite{16}. But
they restrict themselves only in black hole solutions (solution of Einstein
equation) without any form fields. When the form field is turned on, the
case becomes quite different. We have to solve the new higher dimensional
Einstein-Maxwell equations.

Therefore, in this paper, we first study what is the nucleation of a $p$%
-brane with form field, Emphasis is put on the explanation of dimension of
field strength. Next, we analyze the relationship between the M5 brane and
the dielectric one in nucleation case. To calculate the decay width of
dielectric brane we have to find the nucleation of rotating M5 brane.
Because it is still an open question till now, we leave it as a future job.
For further investigation, we write down a note about rotating black holes,
and discuss how to generalize the nucleation to the flux branes.

\section{Form fields in nucleation}

Let us start from the static solutions of $p$-branes. Consider the
Lagrangian that there is a single field strength and a single or no dilaton (%
$a=0$) field \cite{7,8}.%
\begin{equation}
e^{-1}\mathcal{L}=R-\frac{1}{2}(\partial \phi )^{2}-\frac{1}{2n!}e^{a\phi
}F^{2},a^{2}=\Delta -\frac{2d\tilde{d}}{D-2}  \label{2.1}
\end{equation}%
(see ref. \cite{7} for symbols) and the metric can be written in the
following%
\begin{equation}
ds_{D}^{2}=e^{2A}(-dt^{2}+dx_{i}dx_{i}+e^{2f}d\tau
^{2})+e^{2B}(e^{-2f}dr^{2}+r^{2}d\Omega _{n}^{2})  \label{2.2}
\end{equation}%
in which, $i=1,\cdot \cdot \cdot ,d-2$, $n=\tilde{d}+1$, $A$, $B$ and $f$
are functions of $r$, $d=p+1=D-\tilde{d}-2$ represents the dimensions of the
brane part. In fact, the metric we use is the same as in ref. \cite{7},
except that the time is continued to an Euclidean coordinate $\tau $, and
one of the space coordinates is continued to the time $t$. When we consider
the reduction from 11-dimension to the 10-dimension, $\tau $ will be chosen
as the compactification coordinate. the solution is \cite{7}%
\begin{equation}
e^{2f}=1-\frac{k}{r^{\tilde{d}}},e^{2A}=H^{-\frac{4\tilde{d}}{\Delta (D-2)}%
},e^{2B}=H^{\frac{4d}{\Delta (D-2)}},H=1+\frac{k}{r^{\tilde{d}}}\sinh
^{2}\mu   \label{2.3}
\end{equation}%
and%
\begin{equation}
F^{M}=\frac{2\tilde{d}k}{\sqrt{\Delta }}\sinh \delta _{m}\cosh \delta
_{m}\epsilon _{\tilde{d}+1}  \label{2.4}
\end{equation}%
for magnetic charged $p$-brane,%
\begin{equation}
F^{E}=\frac{2\tilde{d}k}{\sqrt{\Delta }}\sinh \delta _{e}\cosh \delta
_{e}\epsilon _{\tilde{d}+1}  \label{2.5}
\end{equation}%
for electric charged $p$-brane.

To nucleate the $p$-brane, we suppose that%
\begin{equation}
ds_{D}^{2}=e^{2A}(dx_{i}dx_{i}+e^{2f}d\tau
^{2})+e^{2B}[e^{-2f}dr^{2}+r^{2}(d\alpha ^{2}+\cos ^{2}\alpha d\Omega
_{n}^{2})]  \label{2.6}
\end{equation}%
in which, $i=1,\cdot \cdot \cdot ,d-1$, $n=\tilde{d}$ and it has a
solution in same form like e.qs. (\ref{2.3}), (\ref{2.4}) and
(\ref{2.5}), but now $d$ and $\tilde{d}$ is decreased and
increased by one respectively. Two points are remarkable: 1) The
nucleation solution is an Euclidean brane, its Lagrangian
evolution is obtained by setting $\alpha =it$ in the transverse
space \cite{26}. 2) We have one order higher magnetic field
strength or one order lower electric field strength. This needs to
be interpreted.

For 10 dimensional theory above mentioned, nucleation solution of $p$-brane
is equivalent to a double analytic continuation of $p-1$ brane. However in
M-theory it is questionable even for continuation, since we do not have M1
and M4 branes. Therefore how to understand these field strengthes,
especially $F_{5}$ (for M5) and $F_{3}$ (for M2) appear in the theory is the
key to the question.

Let us explain it by using an example of D6 brane carrying a magnetic
monopole which is reduced from 11-dimensional Taub NUT solution (For short,
we consider the extreme case i.e. $f=0$. And the following is in Einstein
frame).%
\begin{equation}
ds_{E}^{2}=H^{-1/8}(-dt^{2}+dx_{i}dx_{i})+H^{7/8}(dr^{2}+r^{2}d\Omega
_{2}^{2}),H=1+\frac{Q_{m}}{r}  \label{2.7}
\end{equation}%
\ in Wu-Yang gauge%
\begin{equation}
F_{2}=dA=\ast dH,A=Q_{m}(\pm 1-\cos \theta )d\varphi  \label{2.8}
\end{equation}%
in which the Hoge star is defined with respect to the flat transverse space.
The corresponding nucleation solution is%
\begin{equation}
ds_{E}^{2}=\mathcal{H}^{-1/4}(dx_{i}dx_{i})+\mathcal{H}^{3/4}[dr^{2}+r^{2}(d%
\alpha ^{2}+\cos ^{2}\alpha d\Omega _{2}^{2})],\mathcal{H}=1+\frac{Q_{m}}{%
r^{2}}  \label{2.9}
\end{equation}%
and%
\begin{equation}
\mathcal{F}_{3}=\ast d\mathcal{H}=Q_{m}\cos ^{2}\alpha \sin \theta d\alpha
d\theta d\varphi  \label{2.10}
\end{equation}%
setting $h(\alpha )=\int \cos ^{2}\alpha d\alpha $, and choosing the
integration constant so that $h(0)=1$, then we can see%
\begin{equation}
\mathcal{F}_{2}(\alpha )=h(\alpha )F_{2}  \label{2.11}
\end{equation}%
\bigskip as the evolution monopole field. Hence%
\begin{equation}
\mathcal{F}_{3}=\partial _{\alpha }\mathcal{F}(\alpha )d\alpha  \label{2.12}
\end{equation}%
\bigskip represents the evolution rate of monopole field by the analytic
continuation $\alpha =it$. As for the ''electric'' brane (e.g. Do), the form
field will be the dual of evolution rate of dual magnetic field.

\section{Dielectric effect}

Now let us study M5 brane in more detail. Because that there is no dilaton
in M-theory%
\begin{equation}
\Delta =\frac{2d\tilde{d}}{D-2},e^{2A}=H^{-2/d},e^{2B}=H^{-2/\tilde{d}}
\label{3.1}
\end{equation}%
so the metric is%
\begin{equation}
ds_{11}^{2}=H^{-1/3}(-dt^{2}+dx_{i}dx_{i}+fd\tau ^{2})+H^{2/3}[\frac{dr^{2}}{%
f}+r^{2}(d\theta ^{2}+\sin ^{2}\theta d\tilde{\varphi}^{2}+\cos ^{2}\theta
d\Omega _{2}^{2})]  \label{3.2}
\end{equation}%
in which%
\begin{equation}
f=1-\frac{2m}{r^{3}},H=1+\frac{2m}{r^{3}}\sinh ^{2}\delta  \label{3.3}
\end{equation}%
and%
\begin{equation}
F_{4}=3(2m)\sinh \delta \cosh \delta \cos ^{2}\theta \sin \theta d\theta d%
\tilde{\varphi}\epsilon (S^{2})  \label{3.4}
\end{equation}

The corresponding nucleation of M5 brane is%
\begin{equation}
ds_{11}^{2}=\mathcal{H}^{-2/5}(dx_{i}dx_{i}+\emph{f}d\tau ^{2})+\mathcal{H}%
^{1/2}\{\frac{dr^{2}}{\emph{f}}+r^{2}[d\alpha ^{2}+\cos ^{2}\alpha (d\theta
^{2}+\sin ^{2}\theta d\tilde{\varphi}^{2}+\cos ^{2}\theta d\Omega
_{2}^{2})]\}  \label{3.5}
\end{equation}%
where%
\begin{equation}
\emph{f}=1-\frac{2m}{r^{4}},\mathcal{H}=1+\frac{2m}{r^{4}}\sinh ^{2}\delta
\label{3.6}
\end{equation}%
and%
\begin{equation}
\mathcal{F}_{5}=d\mathcal{A}_{4},\mathcal{A}_{4}=\frac{12}{\sqrt{10}}2m\sinh
\delta \cosh \delta \cos ^{4}\alpha \cos ^{3}\theta d\alpha d\tilde{\varphi}%
\epsilon (S^{2})  \label{3.7}
\end{equation}%
can be also interpreted as the evolution rate of the evolution version of $%
F_{4}$. For the nucleation slice $\alpha =0$, the metric has the same
topology as the original M5 brane. The subsequent Lagrangian evolution is
obtained by setting $\alpha =it$.

As mentioned in introduction, Cveti\v{c} and Youm generalized the Kerr-like
black hole to the rotating M5 brane \cite{10,11}. It seems that one might
find a nucleation version of this stabilized solution. However, it is a pity
that we have not get the correct rotating solution for nucleation M5 brane
till now. Therefore we would like to mimic ref. \cite{21} to analyze the
so-called dielectric effect of nucleation M5 brane in ''maximal magnetic
field'' case \cite{16}.

First of all we give the compactification radius by using of missing conical
singularity condition%
\begin{equation}
R_{11}=g\sqrt{\alpha ^{%
{\acute{}}%
}}=\frac{r_{H}}{2}(\cosh \delta )^{9/10},r_{H}^{4}=2m  \label{3.8}
\end{equation}%
Then we perform the reduction along the Killing vector field%
\begin{equation}
q=\frac{\partial }{\partial \tau }+B\frac{\partial }{\partial \tilde{\varphi}%
}  \label{3.9}
\end{equation}%
setting $\varphi =\tilde{\varphi}+B\tau $, the IIA metric can be written as%
\begin{eqnarray}
ds_{10}^{2} &=&\Lambda ^{1/2}\mathcal{H}^{-2/5}ds^{2}(\mathbb{E}%
^{4})+\Lambda ^{1/2}\mathcal{H}^{1/2}\{\frac{dr^{2}}{\emph{f}}+r^{2}[d\alpha
^{2}+\cos ^{2}\alpha (d\theta ^{2}+\cos ^{2}\theta d\Omega _{2}^{2})]\}
\label{3.10} \\
&&+\Lambda ^{-1/2}\mathcal{H}^{1/10}\emph{f}r^{2}\cos ^{2}\alpha \sin
^{2}\theta d\varphi ^{2}  \notag
\end{eqnarray}%
in which%
\begin{equation}
\Lambda =\mathcal{H}^{-2/5}(\emph{f}+\mathcal{H}^{9/10}B^{2}r^{2}\cos
^{2}\alpha \sin ^{2}\theta )=e^{4\phi /3}  \label{3.11}
\end{equation}%
To avoid the conical singularity, set the magnetic field parameter%
\begin{equation}
B=1/R_{11}  \label{3.12}
\end{equation}%
\ \ \ \ \ And we have the following form fields%
\begin{eqnarray}
\mathcal{A}_{\varphi } &=&\frac{1}{\Lambda }\mathcal{H}^{1/2}\cos ^{2}\alpha
\sin ^{2}\theta Bd\varphi ,  \label{3.13} \\
\mathcal{A}_{3} &=&\frac{12B}{\sqrt{10}}2m\sinh \delta \cosh \delta \cos
^{4}\alpha \cos ^{3}\theta d\alpha \epsilon (S^{2}),  \notag \\
\mathcal{A}_{4} &=&\frac{12}{\sqrt{10}}2m\sinh \delta \cosh \delta \cos
^{4}\alpha \cos ^{3}\theta d\alpha d\varphi \epsilon (S^{2})  \notag
\end{eqnarray}%
for $r\gg r_{H}$ (and $\sinh ^{2}\delta $ is not too big) the string metric
is reduced to%
\begin{eqnarray}
ds_{10}^{2} &=&\Lambda ^{1/2}\{ds^{2}(\mathbb{E}^{4})+dr^{2}+r^{2}[d\alpha
^{2}+\cos ^{2}\alpha (d\theta ^{2}+\cos ^{2}\theta d\Omega _{2}^{2})]\}
\label{3.14} \\
&&+\Lambda ^{-1/2}r^{2}\cos ^{2}\alpha \sin ^{2}\theta d\varphi ^{2}  \notag
\end{eqnarray}%
using the coordinate transformation%
\begin{eqnarray}
\rho &=&r\cos \alpha \sin \theta ,  \label{3.15} \\
\upsilon \cos \chi &=&r\cos \alpha \cos \theta ,  \notag \\
\upsilon \sin \chi &=&r\sin \alpha  \notag
\end{eqnarray}%
the metric becomes%
\begin{equation}
ds_{10}^{2}=\Lambda ^{1/2}[ds^{2}(\mathbb{E}^{4})+ds^{2}(\mathbb{M}%
^{4})+d\rho ^{2}]+\Lambda ^{-1/2}\rho ^{2}d\varphi ^{2}  \label{3.16}
\end{equation}%
in which we made a replacement of $d\upsilon ^{2}+\upsilon ^{2}(d\chi
^{2}+\cos ^{2}\chi d\Omega _{2}^{2})\rightarrow ds^{2}(\mathbb{M}^{4})$ when
one of the coordinate is analytic continued to the time. The dual of the $2$
form RR field strength is%
\begin{equation}
\mathcal{F}_{8}=2B\epsilon (\mathbb{E}^{4})\wedge \epsilon (\mathbb{M}^{4})
\label{3.17}
\end{equation}%
represents a flux $7$ brane $\mathbf{F}_{7}$.

The case we have studied corresponds to a dielectric brane immersed in a $%
\mathbf{F}_{7}$ brane, in which D4 branes dissolve in a D6 brane. So this is
the nucleation of the dielectric brane. Of course, we may also consider the
near core approximation: $r^{2}=r_{H}^{2}+\tilde{r}^{2}$, $\tilde{r}$, $%
\theta \approx 0$. we have%
\begin{equation}
\emph{f}\simeq 2\frac{\tilde{r}^{2}}{r_{H}^{2}}=4\lambda ^{2},\mathcal{H}%
\simeq \cosh ^{2}\delta ,\Lambda \simeq 4(\cosh \delta )^{-4/5}(\lambda
^{2}+\cos ^{2}\alpha \theta ^{2})  \label{3.18}
\end{equation}%
and%
\begin{eqnarray}
ds_{10}^{2} &\simeq &\sqrt{\lambda ^{2}+\cos ^{2}\alpha \theta ^{2}}(\cosh
\delta )^{-6/5}[ds^{2}(\mathbb{E}^{4})+(\cosh \delta
)^{9/5}r_{H}^{2}[d\lambda ^{2}  \label{3.19} \\
&&+d\alpha ^{2}+\cos ^{2}\alpha (d\theta ^{2}+d\Omega _{2}^{2}+\frac{\lambda
^{2}\theta ^{2}}{\lambda ^{2}+\cos ^{2}\alpha \theta ^{2}}d\varphi ^{2})]
\notag
\end{eqnarray}%
setting%
\begin{equation}
\lambda ^{2}+\theta ^{2}\cos ^{2}\alpha =(\cosh \delta )^{3/10}\frac{\rho }{%
r_{H}},\lambda \theta \cos \alpha =\frac{1}{2}(\cosh \delta )^{3/10}\frac{%
\rho }{r_{H}}\sin \eta   \label{3.20}
\end{equation}%
and%
\begin{equation}
\mu =\frac{r_{H}}{4}(\cosh \delta )^{21/10}  \label{3.21}
\end{equation}%
in $\alpha \approx 0$ neighborhood we obtain%
\begin{equation}
ds_{10}^{2}=\sqrt{\frac{\rho }{\mu }}ds^{2}(\mathbb{E}^{4})+(\cosh \delta
)^{9/5}\sqrt{\frac{\rho }{\mu }}r_{H}^{2}(d\alpha ^{2}+\cos ^{2}\alpha
d\Omega _{2}^{2})+\sqrt{\frac{\mu }{\rho }}ds^{2}(\mathbb{E}^{3})
\label{3.22}
\end{equation}%
where we use $d\rho ^{2}+\rho ^{2}(d\eta +\sin ^{2}\eta d\varphi
^{2})\rightarrow ds^{2}(\mathbb{E}^{3})$ and the form fields become%
\begin{eqnarray}
\mathcal{A}_{\varphi } &=&(\cosh \delta )^{-6/5}\mu (1-\cos \eta )d\varphi ,
\label{3.23} \\
\mathcal{A}_{3} &=&\frac{9}{\sqrt{10}}r_{H}^{3}\sinh 2\delta \cos
{}^{2}\alpha \lbrack (\cosh \delta )^{-3/5}\rho (1-\cos \eta )-\frac{4}{3}%
(\cosh \delta )^{-9/10}\cos ^{2}\alpha ]d\alpha \epsilon (S^{2}),  \notag \\
\mathcal{A}_{4} &=&\frac{9}{\sqrt{10}}r_{H}^{4}\sinh 2\delta \cos
{}^{2}\alpha \lbrack \frac{1}{2}(\cosh \delta )^{3/10}\rho (1-\cos \eta )-%
\frac{2}{3}\cos ^{2}\alpha ]d\alpha d\varphi \epsilon (S^{2})  \notag
\end{eqnarray}

\section{About the nucleation of rotating black holes}

One question for preparing the nucleation of rotating M5 brane we would like
to state is that instead of analytically continuing in one of the ignorable
angles in e.q.(5.2) of ref. \cite{16} we suggest to use formula%
\begin{eqnarray}
ds^{2} &=&f(d\tau -\frac{2ml}{r^{p}\rho ^{2}f}\cos ^{2}\alpha \sin
^{2}\theta d\varphi )^{2}+\frac{dr^{2}}{\tilde{f}}-l^{2}(\cos \alpha \cos
\theta d\theta -\sin \alpha \sin \theta d\alpha )^{2}  \label{4.1} \\
&&+r^{2}[d\alpha ^{2}+\cos ^{2}\alpha (d\theta ^{2}+\cos ^{2}\theta d\Omega
_{p}^{2})]+\frac{1}{f}(r^{2}-l^{2}-\frac{2m}{r^{p}})\cos ^{2}\alpha \sin
^{2}\theta d\varphi ^{2}  \notag
\end{eqnarray}%
where%
\begin{equation}
f=1-\frac{2m}{r^{p}\rho ^{2}},\tilde{f}=f-\frac{l^{2}\cos ^{2}\alpha \sin
^{2}\theta }{\rho ^{2}},\rho ^{2}=r^{2}-l^{2}+l^{2}\cos ^{2}\alpha \sin
^{2}\theta  \label{4.2}
\end{equation}%
Although these two formula are equivalent with each other in
mathematics, They are not equivalent in physical meaning. Because
that in e.q.(\ref{4.1}) the evolution space is whole fixed points
of killing vector, But it is only a subspace of them in the former
case.

\section{Nucleation of flux brane}

Our proposition about nucleation of $p$-brane is also applicable to flux
branes. It is known that $\mathbf{F}_{p}$ in type II strings, $\mathbf{F}%
_{3} $\ and $\mathbf{F}_{6}$\ in M-theory have been calculated in refs \cite%
{19,20,21}. We can use double analytic continuation to find the nucleation
of $\mathbf{F}_{p}$\ brane from $\mathbf{F}_{p-1}$\ brane for type II flux
branes just like above mentioned general type II $p$-branes. For flux branes
in M-theory like the M5 brane we may get nucleation $\mathbf{F}_{6}$ brane
solution as follows by using same symbols and same method in ref. \cite{21}.
Now $D=11$, $d=6$, $\tilde{d}=4$,%
\begin{equation}
f^{"}=-\frac{20}{9}E^{2}e^{f}+24E^{2}e^{g},g^{"}=E^{2}e^{f}+18E^{2}e^{g}
\label{5.1}
\end{equation}%
setting $e^{g}=\eta e^{f}$ , so that $\eta =\frac{29}{54}$,and%
\begin{equation}
e^{f}=\frac{3}{(4Er)^{2}},e^{2A}=\eta ^{1/6}e^{f/24},e^{2C}=\eta
e^{5f/4},e^{2B}=e^{f/4}  \label{5.2}
\end{equation}%
Then the metric for nucleation of $\mathbf{F}_{6}$ brane in terms of the
proper radial distance coordinate $u$ is%
\begin{equation}
ds^{2}=(\frac{E^{2}u^{2}}{3})^{1/6}ds^{2}(\mathbb{E}^{6})+du^{2}+\frac{18}{29%
}u^{2}(d\alpha ^{2}+\cos ^{2}\alpha d\Omega _{3}^{2})  \label{5.3}
\end{equation}%
with%
\begin{equation}
\mathcal{F}_{6}=E\epsilon (\mathbb{E}^{6})  \label{5.4}
\end{equation}%
which can be identified as the dual of the evolution rate of evolution
version of%
\begin{equation}
\mathcal{\tilde{F}}=\frac{1}{E^{2}}(\frac{2}{7E^{2}r^{2}})^{4/3}dr\epsilon
(S^{3})=(\frac{4^{3}7^{4}}{3^{5}})^{1/2}\frac{u^{4}}{E^{2}}du\epsilon (S^{3})
\label{5.5}
\end{equation}%
Similarity, the nucleation of flux brane $\mathbf{F}_{3}$ can also be
written.

\section{Prospect}

We expect to gain the nucleation solution for rotating M5 brane. Then we can
find the values of the radius of the relevant sphere and calculate the
Euclidean action of the rotating system. Finally we can obtain the decay
width of the dielectric brane.

\bigskip

\textbf{Acknowledgement:} Our work is supported by National Natural Science
Foundation of China.

\end{document}